\documentclass[longauth,a4]{aa}

\usepackage{graphicx}
\usepackage{txfonts}

\newcommand{\etal}{\hbox{et al.} } 
\newcommand{\micro}{\mbox{\usefont{U}{eur}{m}{n}\char22}}

\begin{document}

\title{\bf Discovery of very high energy gamma-ray emission coincident with molecular clouds in the \object{W~28} (G6.4$-$0.1) field}

\titlerunning{Associated VHE and CO emission in the W~28 field}

\date{Received / Accepted}

\offprints{rowell@physics.adelaide.edu.au, mori@aserv.a.phys.nagoya-u.ac.jp}

\author{F. Aharonian\inst{1,13}
 \and A.G.~Akhperjanian \inst{2}
 \and A.R.~Bazer-Bachi \inst{3}
 \and B.~Behera \inst{14}
 \and M.~Beilicke \inst{4}
 \and W.~Benbow \inst{1}
 \and D.~Berge \inst{1} \thanks{now at CERN, Geneva, Switzerland}
 \and K.~Bernl\"ohr \inst{1,5}
 \and C.~Boisson \inst{6}
 \and O.~Bolz \inst{1}
% ...
 \\[3mm] \mailname{growell@physics.adelaide.edu.au, mori@aserv.a.phys.nagoya-u.ac.jp}
}

\institute{
Max-Planck-Institut f\"ur Kernphysik, P.O. Box 103980, D 69029
Heidelberg, Germany
\and
 Yerevan Physics Institute, 2 Alikhanian Brothers St., 375036 Yerevan,
Armenia
\and
Centre d'Etude Spatiale des Rayonnements, CNRS/UPS, 9 av. du Colonel Roche, BP
4346, F-31029 Toulouse Cedex 4, France
\and
Universit\"at Hamburg, Institut f\"ur Experimentalphysik, Luruper Chaussee
149, D 22761 Hamburg, Germany
\and
Institut f\"ur Physik, Humboldt-Universit\"at zu Berlin, Newtonstr. 15,
D 12489 Berlin, Germany
\and
LUTH, UMR 8102 du CNRS, Observatoire de Paris, Section de Meudon, F-92195 Meudon Cedex,
France
\and
DAPNIA/DSM/CEA, CE Saclay, F-91191
Gif-sur-Yvette, Cedex, France
\and
University of Durham, Department of Physics, South Road, Durham DH1 3LE,
U.K.
\and
Unit for Space Physics, North-West University, Potchefstroom 2520,
    South Africa
\and
Laboratoire Leprince-Ringuet, IN2P3/CNRS,
Ecole Polytechnique, F-91128 Palaiseau, France
\and 
Laboratoire d'Annecy-le-Vieux de Physique des Particules, IN2P3/CNRS,
9 Chemin de Bellevue - BP 110 F-74941 Annecy-le-Vieux Cedex, France
\and
APC, 11 Place Marcelin Berthelot, F-75231 Paris Cedex 05, France 
\thanks{UMR 7164 (CNRS, Universit\'e Paris VII, CEA, Observatoire de Paris)}
\and
Dublin Institute for Advanced Studies, 5 Merrion Square, Dublin 2,
Ireland
\and
Landessternwarte, Universit\"at Heidelberg, K\"onigstuhl, D 69117 Heidelberg, Germany
% ...
}

\authorrunning{Aharonian \etal}

\date{Received / Accepted}


\abstract
{}
{ Observations of shell-type supernova remnants (SNRs) in the GeV to multi-TeV $\gamma$-ray band, coupled with those
  at millimetre radio wavelengths, are motivated by the
  search for cosmic-ray accelerators in our Galaxy. The old-age mixed-morphology SNR W~28 (distance $\sim$2~kpc) is a prime target 
  due to its interaction with molecular clouds along its northeastern boundary and other clouds situated nearby.} 
{ We observed the W~28 field (for $\sim$40~h) at very high energy (VHE) $\gamma$-ray energies ($E>$0.1~TeV) with the H.E.S.S. 
  Cherenkov telescopes. A reanalysis of EGRET $E>$100~MeV data was also undertaken.
  Results from the NANTEN 4m telescope Galactic plane survey and other CO observations were used to 
  study molecular clouds. 
}
{ We have discovered VHE $\gamma$-ray emission (HESS~J1801$-$233) coincident with the northeastern boundary of W~28 
  and a complex of sources (HESS~J1800$-$240A, B and C) $\sim$0.5$^\circ$ south of W~28 in the Galactic disc. The EGRET source
  (GRO~J1801$-$2320) is centred on HESS~J1801$-$233 but may also be related to HESS~J1800$-$240 given the
  large EGRET point spread function. 
  The VHE differential photon spectra are well fit by pure power laws with indices $\Gamma \sim 2.3$ to 2.7. The spectral
  indices of HESS~J1800$-$240A, B, and C are consistent within statistical errors.
  All VHE sources are $\sim$10$^\prime$ in intrinsic radius except for HESS~J1800$-$240C, which appears pointlike. 
  The NANTEN $^{12}$CO(J=1-0) data reveal molecular clouds positionally associating with the 
  VHE emission, spanning a $\sim$15~km~s$^{-1}$ range in local standard of rest velocity.
}
{ 
  The VHE/molecular cloud association could indicate a hadronic origin for HESS~J1801$-$233 and HESS~J1800$-$240,
  and several cloud components in projection may contribute to the VHE emission. 
  The clouds have components covering a broad velocity range encompassing the distance estimates for 
  W~28 ($\sim$2~kpc) and extending up to $\sim$4~kpc. Assuming hadronic origin and distances 
  of 2 and 4~kpc for cloud components, the required cosmic-ray density enhancement factors (with respect to the solar
  value) are in the range $\sim$10 to $\sim$30. If situated at 2~kpc distance, 
  such cosmic-ray densities may be supplied by SNRs like W~28. Additionally and/or alternatively, particle acceleration 
  may come from several catalogued SNRs and SNR candidates, the energetic ultra compact HII region W~28A2, and the HII regions 
  M~8 and M~20, along with their associated open clusters. 
  Further sub-mm observations would be recommended to
  probe in detail the dynamics of the molecular clouds at velocites $>$10~km~s$^{-1}$ and their possible connection to W~28.
}
\maketitle

\section{Introduction: W~28 and surroundings}

The study of shell-type supernova remnants (SNRs) at $\gamma$-ray energies is motivated
by the long-held idea that they are the dominant sites of hadronic Galactic cosmic-ray (CR)
acceleration to energies approaching the \emph{knee} ($\sim 10^{15}$~eV) (e.g. Ginzburg \& Syrovatskii \cite{Ginzburg:1}, 
Blandford \& Eichler \cite{Blandford:1}). CRs (hadrons and electrons) are injected into the SNR shock front, 
and are then accelerated via the diffusive shock acceleration (DSA) process 
(for a review see Drury \cite{Drury:2}). 
Subsequent $\gamma$-ray production from the interaction of these CRs with ambient 
matter and/or electromagnetic fields is a tracer of such non-thermal particle acceleration, 
and establishing the hadronic or electronic nature of the parent CRs in any $\gamma$-ray source remains a key
issue. 
Two SNRs, RX~J1713.7$-$3946 and RX~J0852.0$-$4622, have so far established shell-like morphology in VHE $\gamma$-rays 
(Aharonian \etal \cite{HESS_RXJ1713,HESS_VelaJnr,HESS_RXJ1713_II,HESS_VelaJnr_II,HESS_RXJ1713_III}), with spectra extending 
to 20~TeV and beyond.
In particular for RX~J1713.7$-$3946, particle acceleration up to at least 100~TeV is inferred from the H.E.S.S. observations.  
Although a hadronic origin of the VHE $\gamma$-ray emission is highly likely in the above cases 
(Aharonian \etal \cite{HESS_RXJ1713_II,HESS_VelaJnr_II}, Berezhko \& V\"olk \cite{Berezhko:1}, Berezhko, 
P\"uhlhofer \& V\"olk \cite{Berezhko:2}), an electronic origin is not ruled out.

Disentangling the electronic and hadronic components in TeV SNRs may be made easier
by studying: (1) SNR $\gamma$-ray spectra well beyond $\sim$10 TeV, an energy regime where electrons suffer strong 
radiative energy losses and due to Klein-Nishina effects the resulting inverse-Compton spectra tend to show a cut-off; 
(2) older SNRs (age approaching 10$^5$ yr) in which accelerated electrons have lost much of their
energy through radiative cooling and do not reach multi-TeV energies; (3) SNRs interacting with adjacent molecular clouds 
of very high densities $n> 10^3$~cm$^{-3}$. 
It is the latter regard especially (and to a certain degree the second) 
which makes the SNR W~28 (G6.4$-$0.1) an attractive target for VHE $\gamma$-ray studies. 
In this paper we outline the discovery of VHE $\gamma$-ray emission from several sites in the W~28 field 
and briefly discuss their relationship with molecular clouds, W~28, and other potential particle accelerators in the region.

W~28 (G6.4$-$0.1) is a mixed-morphology
SNR, with dimensions 50$^\prime$x45$^\prime$ and an estimated distance between 1.8 and 3.3~kpc 
(eg. Goudis \cite{Goudis:1}, Lozinskaya \cite{Lozinskaya:1}). 
It is an old-age SNR (age 35000 to 150000~yr; eg. Kaspi \etal \cite{Kaspi:1}), thought to have entered its 
radiative phase of evolution (eg. Lozinskaya \cite{Lozinskaya:1}) in which much of its CRs have escaped into the surrounding 
interstellar medium (ISM). We note also that the evolutionary status (Sedov and/or radiative) of 
shell-type SNRs may depend on the density of their surroundings (see eg. Blondin \etal \cite{Blondin:1}). 

W~28 is distinguished by its interaction with a molecular cloud (Wootten \cite{Wootten:1}) 
along its north and northeastern boundaries. This interaction is traced by the high 
concentration of 1720~MHz OH masers (Frail \etal \cite{Frail:2}, Claussen \etal \cite{Claussen:1,Claussen:2}), 
and also the location of very high-density ($n>10^3$~cm$^{-3}$) shocked gas (Arikawa \etal  \cite{Arikawa:1}, 
Reach \etal \cite{Reach:1}). 
The shell-like radio emission (Long \etal \cite{Long:1}, Dubner \etal \cite{Dubner:1}) peaks at the northern and northeastern
boundaries where interaction with the molecular cloud is established. Further indication of the influence of W~28 on its 
surroundings is the expanding HI void at 
a distance $\sim$1.9~kpc 
(Vel\'azquez \etal \cite{Velazquez:1}). 
The X-ray emission, which overall is well-explained by a thermal model, peaks in the SNR centre but has local enhancements in a region 
overlapping the northeastern SNR/molecular cloud interaction (Rho \& Borkowski \cite{Rho:2}). 

In the neighbourhood of W~28 are the radio-bright HII regions M~20 (Trifid Nebula at $d \sim$1.7~kpc 
Lynds \etal \cite{Lynds:1} -- with open
cluster NGC~6514), M~8 (Lagoon Nebula at $d\sim 2$~kpc Tothill \etal \cite{Tothill:1} --- containing the open clusters 
NGC~6523 and NGC~6530) and the ultra-compact HII region W~28A2, all of which are representative of the massive star 
formation taking place in the region. Further discussion concerning the active star formation in this
region may be found in van den Ancker \etal (\cite{Ancker:1}) and references therein.
Additional SNRs in the vicinity of W~28 have also been identified: G6.67$-$0.42 and G7.06$-$0.12 
(Yusef-Zadeh \etal \cite{Yusef:1}),
G5.55+0.32, G6.10+0.53 and G7.20+0.20 (Brogan \etal \cite{Brogan:1}). 
The pulsar PSR~J1801$-$23
with 
spin-down luminosity $\dot{E} \sim 6.2\times 10^{34}$ erg~s$^{-1}$ and distance $d = 13.5$~kpc (based on its dispersion measure)
is at the northern radio edge (Kaspi \cite{Kaspi:1}). More recent discussion (Claussen \etal \cite{Claussen:3}) 
assigns a lower limit of 9.4$\pm$2.4~kpc for the pulsar distance.

W~28 has also been linked to $\gamma$-ray emission detected at $E>300$~MeV by COS-B (Pollock \cite{Pollock:1}) and $E>100$~MeV 
by EGRET (Sturner \& Dermer \cite{Sturner:1}, Esposito \etal \cite{Esposito:1}, Zhang \etal \cite{Zhang:1}).  
The EGRET source, listed in the 3rd catalogue (Hartman \etal \cite{Hartman:1}) as 3EG~J1800$-$2338, 
is positioned at the southern edge of the radio shell.
We have also performed an analysis of EGRET data, with additional data not included in the 3rd catalogue, and results are discussed later
in this paper.

Previous observations of the W~28 region at VHE energies by the CANGAROO-I telescope revealed no evidence for such emission 
(Rowell \etal \cite{Rowell:1}) and upper limits at the $\sim$0.2 to 0.5 Crab-flux level for energies $E>1.5$~TeV 
(1.1 to 2.9$\times 10^{-11}$~erg~cm$^{-2}$~s$^{-1}$) were set for various regions.

\section{\boldmath Results at VHE and $E>$100~MeV $\gamma$-ray energies}

\subsection{H.E.S.S. VHE analysis and results}
The High Energy Stereoscopic System (H.E.S.S.) was used to observe the W~28 region.
Operating in the Southern Hemisphere, H.E.S.S. consists of four 
identical 13~m diameter Cherenkov telescopes (Bernlohr \etal \cite{Bernlohr:1}). 
H.E.S.S. employs the stereoscopic imaging atmospheric Cherenkov technique, and is sensitive to $\gamma$-rays 
above an energy threshold of $\sim$0.1~TeV (Funk \etal \cite{Funk:1}).  An angular resolution
of 5$^\prime$ to 6$^\prime$ (Gaussian standard deviation) on an event-by-event basis is achieved, and the large field of view (FoV) 
with full width at half maximum $\mathrm{FWHM}\sim 3.5^\circ$ 
permits survey coverage in a single pointing. A point source sensitivity approaching 0.01 Crab flux 
($\sim10^{-13}$~erg~cm$^{-2}$s$^{-1}$ at 1~TeV) is achieved for a 5$\sigma$ detection after $\sim$25~hr observation. 
Further details concerning H.E.S.S. can be found in Hinton (\cite{Hinton:1}) 
and references therein.

The total observation time covering the W~28 region amounts to $\sim$42~hr in a series of runs (with typical duration 
$\sim$28~min) spread over the 2004, 2005 and 2006 seasons.
Runs were accepted for analysis if they 
met quality control criteria based on the recorded rate of isotropic CR background events, 
the number of malfunctioning pixels in each camera, 
the calibration and the tracking performance (see Aharonian \etal \cite{HESS_Calibration} for details).

Data were analysed using the moment-based Hillas analysis procedure, the same used in the analysis of the  
inner Galactic Plane Scan datasets (Aharonian \etal \cite{HESS_GalScan,HESS_GalScan_II}). Observations covered a range
of zenith angles leading to energy thresholds of $\sim 320$~GeV with {\em hard cuts} (Cherenkov image integrated intensity or {\em size} 
$>$200 photoelectrons)
and $\sim$150~GeV for {\em standard cuts} ({\em size} $>$ 80 photoelectrons). {\em Hard cuts} were used in VHE $\gamma$-ray images, 
source location studies and energy spectra. In addition, {\em Standard cuts} were used in energy spectra in order to increase the 
energy coverage of extracted spectra.
Generally consistent results were obtained using an alternative analysis based on a model of Cherenkov image parameters
(de~Naurois \cite{Mathieu:1}), which also utilises an independent calibration and lower cut on image {\em size} of $>$60 
photoelectrons. A forthcoming paper will highlight results in detail from this analysis, which achieves improved sensitivities
at lower thresholds compared to the pure Hillas-based analysis.

\begin{figure*}[t!]
  \sidecaption
  \includegraphics[width=0.75\textwidth]{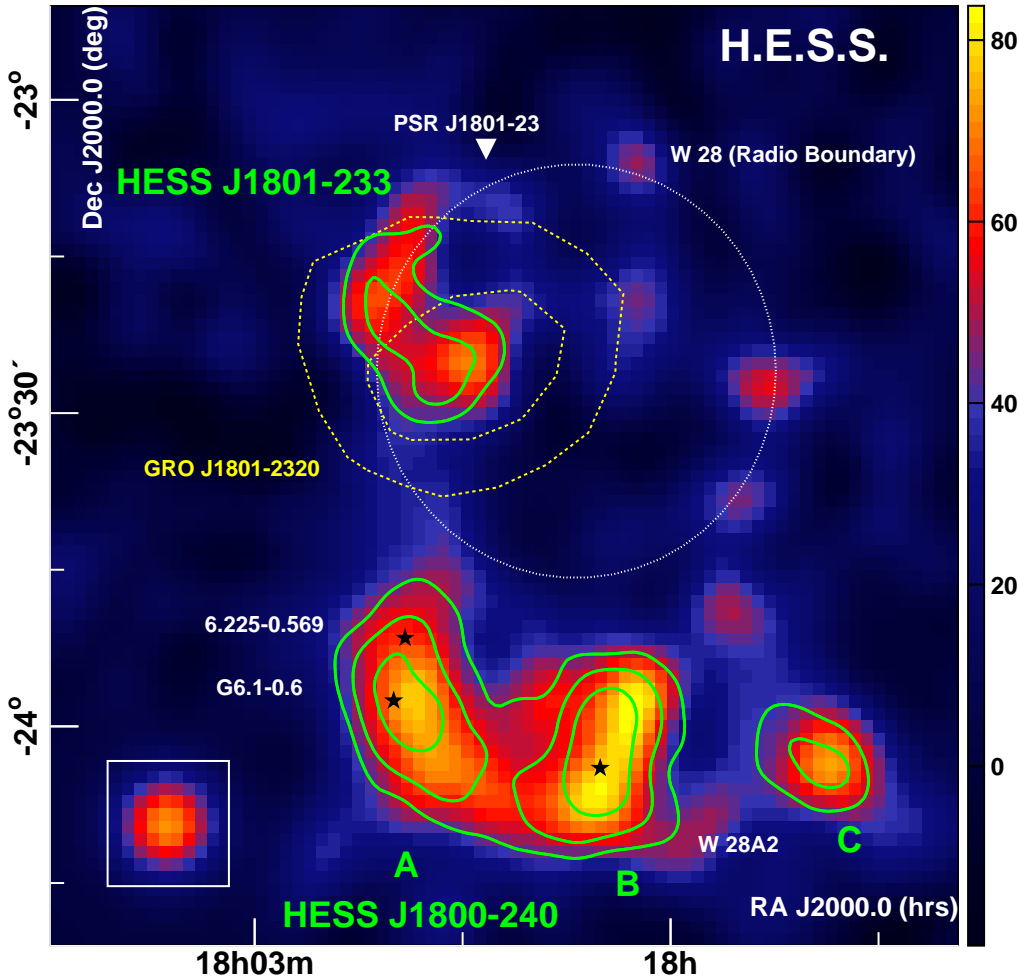}
  \caption{Image (1.5$^\circ\times 1.5^\circ$) of the VHE $\gamma$-ray excess counts (events), corrected for 
    exposure and smoothed with a Gaussian of radius 4.2$^\prime$ (standard deviation). 
    Overlaid are solid green contours of VHE excess 
    (pre-trial) 
    significance levels of 4, 5, and 6$\sigma$, after integrating events within an oversampling radius $\theta$=0.1$^\circ$ 
    appropriate for pointlike sources. 
    The thin-dashed circle depicts the approximate radio boundary of the SNR W~28 guided predominantly by the bright 
    northern emission
    (see Dubner \etal \cite{Dubner:1} \& Brogan \etal
    \cite{Brogan:1}).
    Identified here are VHE source regions HESS~J1801$-$233 to the northeast, 
    and a complex of sources HESS~J1800-240 (A, B \& C) to the south of W~28. 
    Also indicated are: HII regions (black stars); \object{W~28A2} (see text), \object{G6.1$-$0.6} (Kuchar \& Clark \cite{Kuchar:1}), 
    \object{6.225$-$0.569} (Lockman \cite{Lockman:1});
    The 68\% and 95\% location contours (thick-dashed yellow lines) of the $E>100$~MeV EGRET source \object{GRO~J1801$-$2320};
    the pulsar \object{PSR~J1801$-$23} (white triangle). The inset to the bottom left depicts a pointlike source for this
    analysis after the Gaussian smoothing applied to the main image.}
  \label{fig:tevskymap}
\end{figure*}

The VHE $\gamma$-ray image (Fig.~\ref{fig:tevskymap}) reveals two sites of 
VHE $\gamma$-ray emission in the direction of the northeastern and southern boundaries of the W~28 SNR.
The colour scale in this figure depicts the Gaussian-smoothed VHE excess counts above a CR background estimate according to the 
{\em template} model (Rowell \cite{Rowell:2}), along with significance contours obtained after integrating events within a radius
of 0.1$^\circ$ from each bin centre (appropriate for pointlike source searching). Similar images were obtained using alternative
CR background models. A smoothing radius of $4.2^\prime$ was used to sufficiently smooth out random fluctuations in the image.
An assessment of the VHE post-trial significances was made from our original search for marginally extended sources, 
which employed an {\em a priori} integration radius $\theta=0.2^\circ$. Under this scheme we applied 
$\sim 2.2\times 10^5$ trials (a very conservative value applied to these data) accumulated in searching for sources in the inner 
Galactic Plane (as in Aharonian \cite{HESS_GalScan}). The pre-trial significance of the VHE sources, at $\geq +7\sigma$, 
is therefore converted to a post-trial significance of $\geq +5\sigma$. 

Based on the significance contours in Fig.~\ref{fig:tevskymap}, we assign labels to the northeastern source, HESS~J1801$-$233, 
and to the complex of sources to the south, HESS~J1800$-$240, according to their best fit positions (fitting a 2D Gaussian and ellipse 
respectively to the unsmoothed excess map). Three components of HESS~J1800$-$240 are identified, labeled here A, B and C from East 
to West. These components 
represent local peaks $\sim 2\sigma$ above their surrounds. Although not convincingly resolved under this analysis these components 
may comprise separate sources (or at least in part) due to their possible relationship with distinct multiwavelength 
counterparts (discussed later). 

Differential photon energy spectra were extracted from HESS~J1801$-$233 and all three components of HESS~J1800$-$240.  
Spectra were well-fit by pure power laws ($dN/dE = k (E/1 {\rm TeV})^{-\Gamma}$) with photon indices $\Gamma \sim 2.5$ to 2.7
in the energy range $\sim$0.3 to $\sim$5~TeV (see Table~\ref{tab:locations} for results). Spectral fits were obtained using fluxes 
from a combination of {\em hard} and {\em standard} cuts to maximise the energy coverage. 
Spectral analysis employed the
{\em reflected background} model (Berge \etal \cite{Berge:1}), in which control regions reflected through each tracking position
(taking care to avoid known VHE $\gamma$-ray sources) were used to estimate the CR background.  
Within the statistical and systematic errors, the photon indices appear consistent throughout HESS~J1800$-$240.
Except for HESS~J1800$-$240C, all of the VHE sources appear extended with intrinsic radii of $\sim$10$^\prime$.
At a distance of 2~kpc, the VHE source luminosities in the energy range 0.3 to 3~TeV would be 
on the order of $10^{33}$~erg~s$^{-1}$.

\subsection{EGRET $E>100$~MeV analysis and results}
We have also analysed EGRET data for the W~28 region, using CGRO observation cycles (OC) 1 to 6. 
This slightly expands on the dataset of the 3rd EGRET catalogue (using OCs 1 to 4; Hartman \etal \cite{Hartman:1}), 
which revealed the pointlike source, 3EG~J1800$-$2338 ($E>100$~MeV).
Our analysis confirms the presence of a pointlike $E>100$~MeV source in this region, here labeled GRO~J1801$-$2320 
(for $E>100$~MeV). GRO~J1801$-$2320 appears slightly shifted ($\sim$0.2$^\circ$) with respect to 
the 3EG position.   
The 3EG position refers to a $E>$100 MeV determination
based on the diffuse model as of Hunter \etal (\cite{Hunter:1}). Our dedicated
analysis of archival EGRET data comprises different analysis compared to the
3EG catalogue. We first employed the finalised EGRET instrumental responses, 
which were made available by 2001 and are considered mandatory for investigating 
an EGRET source under conditions applicable from the end of OC~4 (narrow field of
view modus; rapidly deteriorating spark chamber efficiency; and other issues). 
Second, we restricted the analysis both in narrowing the data selection to pointing angles with respect
to our region of interest, which avoids the need to invoke a wide-angle point spread function (PSF).
Thirdly, the imprecision of the interstellar emission model was countered via 
adjustments on analysis parameters {\tt gmult} and {\tt gbias} to account for local deviations 
from the large-scale diffuse emission model in the region of interest.
The 68\% and 95\% location contours of GRO~J1801$-$2320 are plotted in Fig.~\ref{fig:tevskymap}, and 
match well the location of HESS~J1801$-$233. Since however the EGRET degree-scale PSF easily encompasses both of the VHE sources,
we cannot rule out a relationship with HESS~J1800$-$240.
For the energy spectrum of GRO~J1801$-$2320, we have used the flux points 
extracted at the position of 3EG~J1800$-$2338 as negligible differences were found between ours and that obtained
at the nominal 3EG position. Fitting a pure power law we obtained a spectral index of $\Gamma = 2.16\pm0.10$,
quite consistent with the published value from Hartman \etal (\cite{Hartman:1}).
Comparisons are made with the VHE spectrum of HESS~J1801$-$233 and HESS~J1800$-240$ in \S\ref{sec:discussion}.
\begin{table*}
  \caption{Numerical summary$^\dagger \, ^\ddagger$ for the VHE and $E>100$~MeV sources in the W~28 region including 
    positional and spectral information.}
  \centering
  \begin{tabular}{llllllll}
    & \multicolumn{2}{c}{Best fit position (J2000.0)} & & & \multicolumn{2}{c}{Spectral analysis} & \\ \cline{2-3} \cline{6-7} \\[-2mm]
   Name & R.A. [deg] & Dec [deg] & $^1$$\sigma_{\rm src}$ [deg] & $^2$$S$ [$\sigma$] (evts) & $^3 k$ & $^4 \Gamma$ & $^5 L$\\ \hline \\[-2mm] 
   HESS~J1801$-$233  & 270.426 $\pm$ 0.031 & $-$23.335 $\pm$ 0.032 & 0.17 $\pm$ 0.03 & +7.9 (281) & 7.50 $\pm$ 1.11 $\pm$ 0.30 & 2.66 $\pm$ 0.27 & 1.5\\
   HESS~J1800$-$240A$^\S$ & 270.491 $\pm$ 0.001 & $-$23.962 $\pm$ 0.001 & 0.15     & +6.0 (180) & 7.65 $\pm$ 1.01 $\pm$ 0.50 & 2.55 $\pm$ 0.18 & 1.5 \\ 
   HESS~J1800$-$240B$^\S$ & 270.110 $\pm$ 0.002 & $-$24.039 $\pm$ 0.009 & 0.15     & +7.8 (236) & 7.58 $\pm$ 0.90 $\pm$ 0.15 & 2.50 $\pm$ 0.17 & 1.4 \\
   HESS~J1800$-$240C & 269.715 $\pm$ 0.014 & $-$24.052 $\pm$ 0.006 & 0.02 $\pm$ 0.15 & +4.5 (71) & 4.59 $\pm$ 0.89 $\pm$ 0.20 & 2.31 $\pm$ 0.35 & 0.8 \\
   HESS~J1800$-$240$^{\S\S}$  & 270.156 $\pm$ 0.044 & $-$23.996 $\pm$ 0.022 & 0.32$^{\rm RA}$ $\pm$ 0.05& +10.3 (652) & 18.63 $\pm$ 1.85 $\pm$ 1.20 & 2.49 $\pm$ 0.14 & 3.6\\  
                                          &                     &                       & 0.17 $^{\rm Dec}$ $\pm$ 0.03 & \\ 
   GRO~J1801$-$2320  & 270.360 $\pm$ 0.150 & $-$23.340 $\pm$ 0.150 &  --  & +13.2        & 3.35 $\pm$ 0.52 & 2.16 $\pm$ 0.10 & 480.0 \\ [1mm] \hline \\[-2mm]
   \multicolumn{8}{l}{\scriptsize  $\dagger$ VHE photon spectra are derived from a region of radius 
     $\theta = \sqrt{0.1^2+\sigma_{\rm src}^2}$ centered on each source's position unless otherwise indicated.}\\
   \multicolumn{8}{l}{\scriptsize  $\ddagger$ In spectra, a function 
     $dN/dE = k E^{-\Gamma}$ ph~cm$^{-2}$~s$^{-1}$~TeV$^{-1}$ is fitted. $E$ is in TeV units (H.E.S.S. data); GeV units (EGRET data).}\\
   \multicolumn{8}{l}{\scriptsize  $1.$ Fitted intrinsic source size (Gaussian std. dev.)}\\
   \multicolumn{8}{l}{\scriptsize  $2.$ Statistical significance and excess events in brackets; for H.E.S.S. sources using Li \& Ma (\cite{Li:1}); for EGRET sources given by $S=\sqrt{T_s}$ for $T_s$ defined by Mattox \etal (\cite{Mattox:1})}\\
   \multicolumn{8}{l}{\scriptsize  $3.$ For H.E.S.S. sources: $\times 10^{-13}$ ph cm$^{-2}$ s$^{-1}$~TeV$^{-1}$ at 1 TeV (with statistical and systematic errors); For EGRET sources:  $\times 10^{-2}$ ph cm$^{-2}$ s$^{-1}$~GeV$^{-1}$ at 1 GeV (with statistical errors)}\\
   \multicolumn{8}{l}{\scriptsize  $4.$ Only statistical errors indicated. Systematic error is estimated at $\pm$0.2}\\
   \multicolumn{8}{l}{\scriptsize  $5.$ Luminosity $\times 10^{33}$ erg s$^{-1}$ at 2~kpc (0.3 to 3 TeV for H.E.S.S.; 0.04 to 6 GeV for EGRET)}\\
   \multicolumn{8}{l}{\scriptsize \S Due to cross contamination between components A \& B, a fixed value of $\sigma_{\rm src}= 0.15^\circ$ estimated visually from Fig.~\ref{fig:tevskymap} was used.}\\
   \multicolumn{8}{l}{\scriptsize \S\S Spectrum extracted from a 0.8$^\circ \times 0.6^\circ$ elliptical region 
     encompassing all components A, B, C, and matching the size of the corresponding molecular cloud.}\\
  \end{tabular}
  \label{tab:locations}
\end{table*}

\section{NANTEN and other observations of Molecular Clouds}
\label{sec:molclouds}

In searching for molecular cloud counterparts to the VHE sources, we analysed
$^{12}$CO ($J$=1--0) molecular line observations taken by the 4-meter mm/sub-mm NANTEN telescope, 
at Las Campanas Observatory, Chile (Mizuno \& Fukui \cite{Mizuno:1}).
The NANTEN Galactic Plane Survey data of 1999 to 2003 (see Matsunaga \etal (\cite{Matsunaga:1}) 
and references therein for details) were used, and for the W~28 region, the survey grid spacing was 4$^{\prime}$. 

\begin{figure*}
  \centering
  \hbox{
    \includegraphics[width=0.5\textwidth]{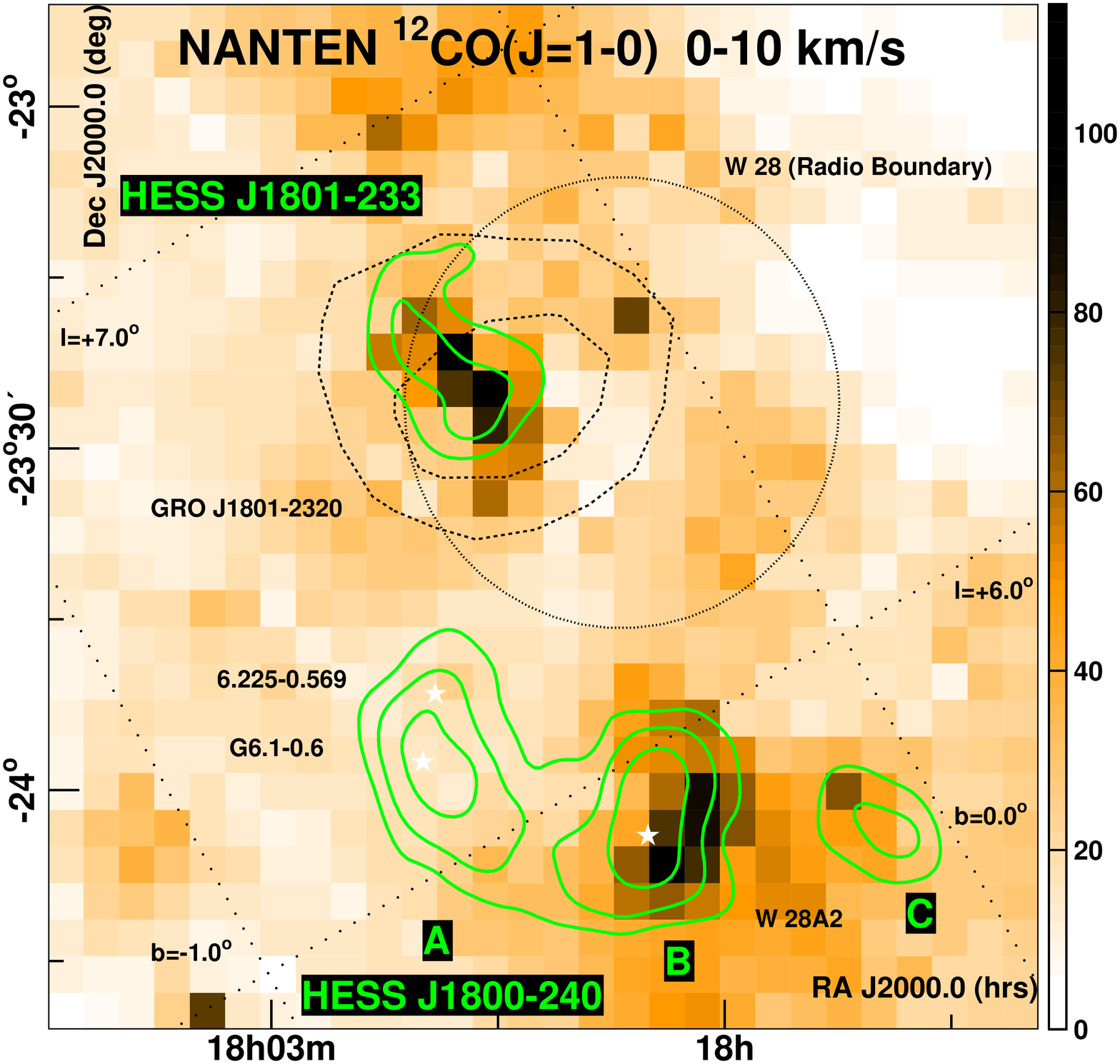}
    \includegraphics[width=0.5\textwidth]{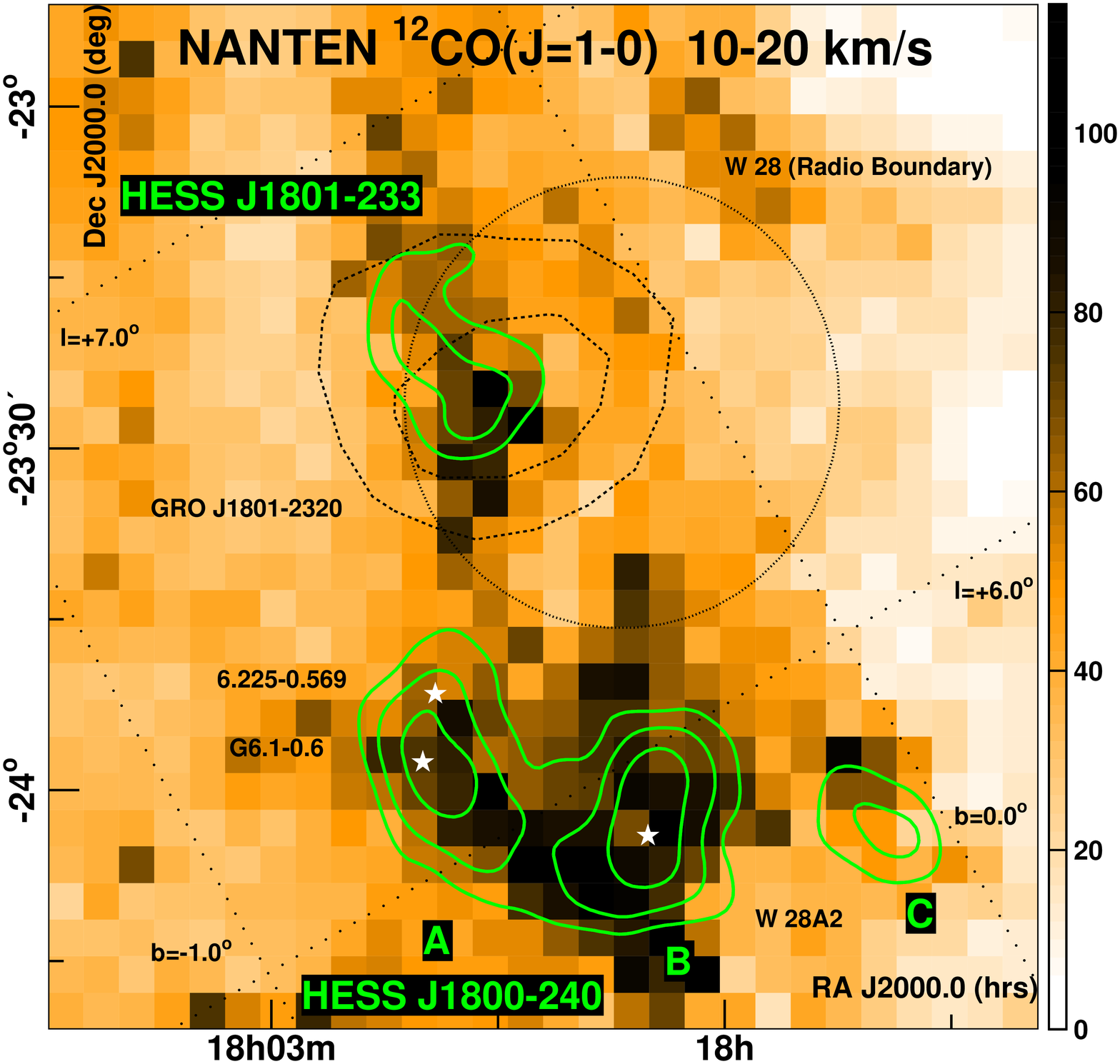}
  }
  \hbox{
    \hspace*{-0.5cm}\includegraphics[width=1.03\textwidth]{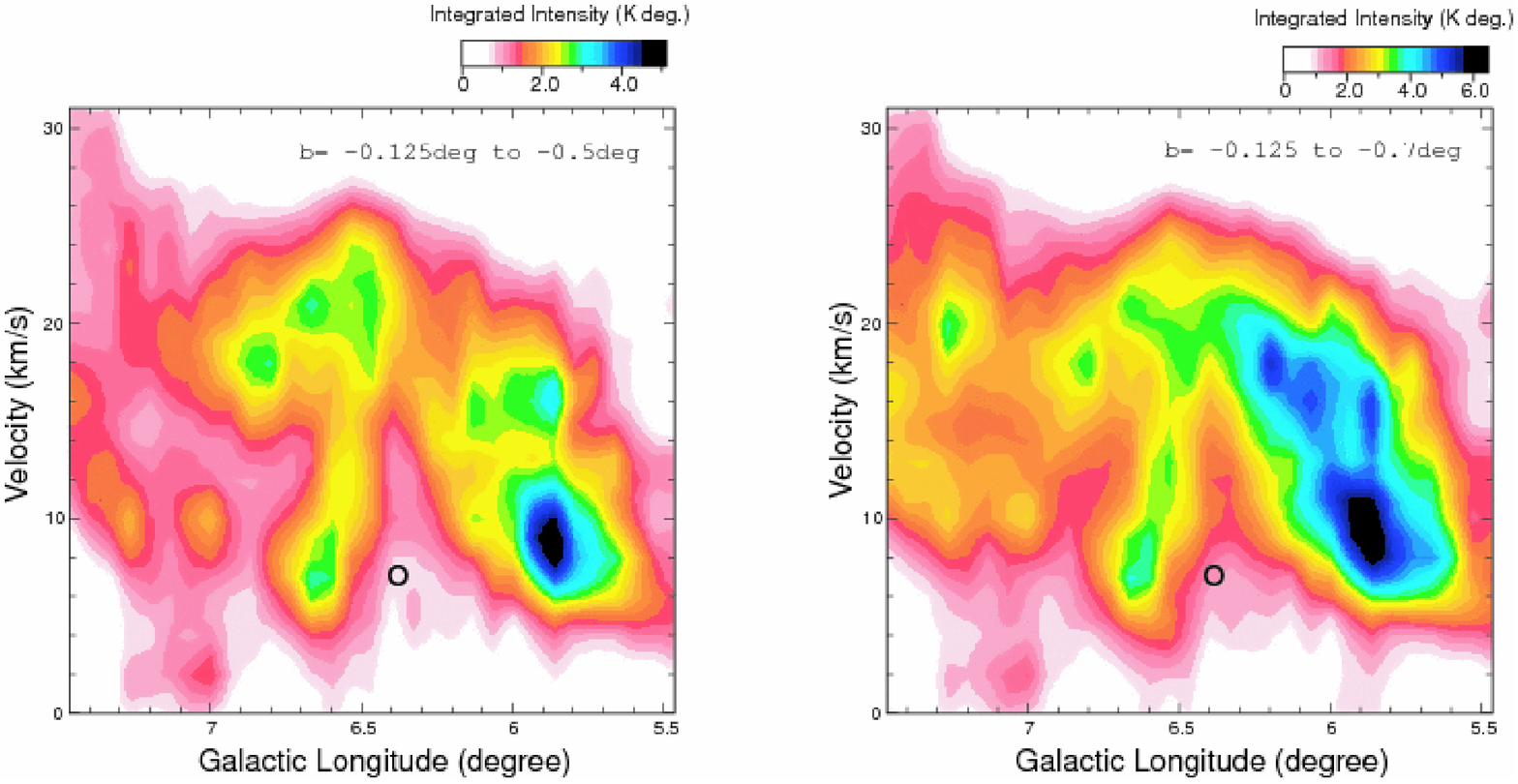}
  }      
  \caption{{\bf Upper Left:} NANTEN $^{12}$CO(J=1-0) image of the W~28 region 
    (linear scale in K~km~s$^{-1}$) for $V_{\rm LSR}$=0 to 10~km~s$^{-1}$ with VHE $\gamma$-ray significance 
    contours overlaid (green) --- levels 4,5,6$\sigma$ as in Fig.~\ref{fig:tevskymap}. 
    The radio boundary of W~28, The 68\% and 95\% location contours of GRO~J1801$-$2320 and the location of the HII region
    W~28A2 (white stars) are indicated. {\bf Upper Right:}  
    NANTEN $^{12}$CO(J=1-0) image for
    $V_{\rm LSR}$=10 to 20~km~s$^{-1}$ (linear scale and same maxima as for upper left panel).
    {\bf Bottom panels:} Distribution of CO emission over the Galactic longitude and $V_{\rm LSR}$ plane integrated over 
    Galactic latitude $b$ ranges $-0.125^\circ$ to $-0.5^\circ$ (left) and  $-0.125^\circ$ to $-0.7^\circ$ (right). The latter range is 
    used to show the effect of extending the latitude range to encompass component A of HESS~J1800$-$240. 
    The bold circle indicates the suggested systemic velocity (7~km~s$^{-1}$) of W~28 from the HI studies 
    of Vel\'azquez \etal (\cite{Velazquez:1}).}
  \label{fig:co_tev}
\end{figure*}

Figure~\ref{fig:co_tev} (upper left panel) shows the $^{12}$CO ($J$=1--0) image integrated over the Local Standard of Rest 
velocity ($V_{\rm LSR}$) range 0 to 10~km~s$^{-1}$, while the right panel shows the image integrated over 
the range $V_{\rm LSR}$=10 to 20~km~s$^{-1}$.
Two prominent $^{12}$CO features representing molecular clouds centred at ($l, b$)=(6.7$^\circ$, $-$0.3$^\circ$) and 
($l, b$)=(5.9$^\circ$, $-$0.4$^\circ$) spatially correspond with the VHE $\gamma$-ray emission. 
As shown in Fig.~\ref{fig:co_tev}, these molecular clouds span 
both $V_{\rm LSR}$ ranges. 
According to the Galactic rotation model of Brand \& Blitz (\cite{Brand:1}), 
these $V_{\rm LSR}$ ranges formally correspond to kinematic distances of 
approximately 0 to $\sim$2.5~kpc (overlapping the Sagittarius arm), and 2.5 to $\sim$4~kpc (reaching the
Scutum-Crux arm) respectively. 
Given the uncertainties in rotation models close to the Galactic centre, such $V_{\rm LSR}$ ranges would cover the distance 
estimates for W~28, the most prominent SNR in the region. Much discussion has centred on the 
systemic velocity (SV) of W~28 (and hence its distance), and how much W~28 has influenced matter in the region. 
H$\alpha$ (Radhakrishman \etal \cite{Radhakrishman:1}) and HI absorption features (Lozinskaya \etal \cite{Lozinskaya:1}) have 
suggested SV$\sim$18~km~s$^{-1}$. Claussen \etal (\cite{Claussen:1}) have pointed to SV$\sim$17~km~$s^{-1}$.
More recent HI studies by Vel\'azquez \etal (\cite{Velazquez:1}) suggest SV=+7~km~s$^{-1}$ 
(which leads to the distance estimate for W~28 at $\sim$1.9~kpc). They also suggest a HI shell 
may also extend over the $V_{\rm LSR}=$ $-$25 to +38~km~s$^{-1}$ range, giving rise to a shock speed of $\sim$30~km~s$^{-1}$. 
Torres \etal (\cite{Torres:1}) and Reach \etal (\cite{Reach:1}) have also studied the large-scale $^{12}$CO(J=1-0) emission for 
this region using the survey data of Dame \etal (\cite{Dame:1}), suggesting that the parent molecular cloud under the influence of W~28
is presently centred at $V_{\rm LSR}\sim$19~km~s$^{-1}$.
The Galactic longitude-velocity (l-v) diagram (bottom panels of Fig.~\ref{fig:co_tev}) from our NANTEN data integrated over the 
Galactic latitude ranges $b=-0.125^\circ$ to $-0.5^\circ$ and $b=-0.125^\circ$ to $-0.7^\circ$ shows the distribution of 
molecular material in relation to the SV of W~28 from the HI studies of Vel\'azquez. The wider, latter $b$ range shows the effect of 
including the cloud component overlapping HESS~J1800$-$240A. 
A void or dip in CO emission appears at a similar $V_{\rm LSR}$ range as found 
in the HI data, with much of the molecular material appearing to surround the void in positive $V_{\rm LSR}$ values 
with respect to the SV of W~28. A similar longitude-velocity picture was revealed by Torres \etal (\cite{Torres:1}) 
(see their Fig.~22). 

The $V_{\rm LSR}$=0 to 10~km~s$^{-1}$ component of the northeast cloud overlapping HESS~J1801$-$233 is already well studied 
(see Reach \etal \cite{Reach:1} and 
references therein). Shocked $^{12}$CO($J$=3--2) molecular gas as indicated by a broad wing-like line dispersion 
(Arikawa \etal \cite{Arikawa:1} --- hereafter A99; using the James Clerk Maxwell Telescope (JCMT); 
in 15$^{\prime\prime}$ grid steps) and a high concentration of OH masers (Claussen \etal \cite{Claussen:1}), 
suggests material here has been compressed by the SNR shock in W~28.
The line dispersion, $\Delta V \leq70$~km~s$^{-1}$, is an indicator of the SNR shock speed in this particular region. 
The unshocked gas was also mapped by A99 via $^{12}$CO($J$=1--0) observations with the Nobeyama 45~m telescope
(in 34$^{\prime \prime}$ grid steps for $V_{\rm LSR}=+4$ to +9~km~s$^{-1}$). 
The shocked and unshocked gas extends to the northeast and northern boundaries of W~28 (see Fig.~3  of A99), and it appears
just their northeastern components are positionally coincident with the VHE emission of HESS~J1801$-$233.
A99 estimate the mass and average density of the shocked gas at  $M\sim 2\times 10^3$~M$_\odot$ and $n\sim10^4$~cm$^{-3}$ 
respectively. For the unshocked gas,
A99 obtained $M\sim 4\times 10^3$~M$_\odot$ and $n\sim 10^3$~cm$^{-3}$ respectively. 
The $V_{\rm LSR}$=10 to 20~km~s$^{-1}$ range in our NANTEN data also reveals additional molecular clouds along the line
of sight that could contribute to the VHE emission.

The southern cloud overlaps all components of HESS~J1800$-$240, with a dominant fraction of the cloud overlapping components A and B. 
The component of this cloud visible in the $V_{\rm LSR}$ = 0 to 10~km~s$^{-1}$
range coincides well with HESS~J1800$-$240B and the HII region W~28A2. 
The strongest CO temperature peak of this component at ($l, b$)=(5.9$^\circ$, -0.4$^\circ$) is within $0.02^\circ$ of W~28A2, 
and is likely the dense material surrounding this HII region.
Moreover the peak's velocity at $V_{\rm LSR}$~9--10 km~s$^{-1}$ (with dispersion of $\sim$15~km~s$^{-1}$),
suggests a distance ($\sim$2.4 kpc) similar to that of W~28A2 ($\sim$2~kpc;  Acord \etal \cite{Acord:1}), and also W~28. 
In the $V_{\rm LSR}$ = 10~km~s$^{-1}$ to 20~km~s$^{-1}$ range, molecular material appears to coincide with 
all three VHE components of HESS~J1800$-$240. In particular, HESS~J1800$-$240A and C have molecular cloud overlaps
only in this latter $V_{\rm LSR}$ range.

Using the relation between the hydrogen column density $N(\rm H_{2})$ and the $^{12}$CO($J$=1--0) intensity (the X-factor)
$W(^{12}$CO), $N({\rm H_2}) = 1.5 \times 10^{20}\ [W({\rm ^{12}CO})/{\rm (K\ km/s)}]\ {\rm (cm^{-2})}$ 
(Strong \etal \cite{Strong:1}), we estimate a total mass for the northeastern cloud from our NANTEN data at 
$\sim 5 \times 10^4$~M$_\odot$ for $d=$2~kpc within an elliptical region of diameter  0.2$^{\circ} \times 0.4^\circ$ 
(7$\times$14~pc; centred on HESS~J1801$-$233) for the velocity range 0--25 km~s$^{-1}$.
An average density (for neutral hydrogen) of $\sim$1.4$\times$10$^{3}$ cm$^{-3}$ is also derived.
Similarly the total mass of the southern cloud is estimated at $\sim$1.0$\times 10^{5}$ $M_{\odot}$
for $d$=2~kpc and combining clouds from a circular area of radius 0.15$^{\circ}$ (5~pc) for the velocity range 12--20 km s$^{-1}$,
and area 0.3$^{\circ} \times 0.6^\circ$ (10.5$\times$21~pc) in diameter for the velocity range 0--12 km~s$^{-1}$ 
(both regions are centred on HESS~J1800$-$240B). The corresponding  
average density is $\sim$1.0$\times$10$^{3}$ cm$^{-3}$.
By integrating over the rather broad 0--20 km~s$^{-1}$ and 0--25 km~s$^{-1}$ ranges we assume that the molecular material 
along this line of sight is physically connected at the same distance (for example $d\sim$2~kpc) and possibly distrupted
or shocked by a local energy source. 
Systematic effects in the mass estimates arise from the velocity crowding in this part of the Galactic plane, and also the
broad velocity range for which X-factor used above may not necessary apply. In the latter case, the X-factor may underestimate
the cloud mass since an appreciable fraction of gas may be heated under the assumption of distrupted and/or shock-heated gas. 
One must allow for $\sim$4~kpc distances for some or even all of the $V_{\rm LSR}>$10~km~s$^{-1}$ cloud components, 
and therefore the conclusion that they are not related to W~28 and other interesting objects at $d\sim 2$~kpc.
If the clouds are related, W~28 could play a disrupting role. The level of this disruption is however unclear since several other
plausible candidates related to the star formation (discussed later) in this region could also contribute. Some other molecular 
cloud complexes have also been discussed as possibly disrupted by adjacent SNRs and/or energetic sources (eg. Yamaguchi \etal 
\cite{Yamaguchi:1}, Moriguchi \etal \cite{Moriguchi:1}).
In table~\ref{tab:masses}, we present a full summary of cloud masses and densities (for regions centered on the VHE 
source coordinates as in table~\ref{tab:locations}) for various combiniations of 
cloud components and distances of 2 and 4~kpc. Velocity separation of cloud components are based on their apparent distribution
in Fig.~\ref{fig:co_tev} (bottom panels).

\section{Radio to X-ray views}

\begin{figure*}[t!]
  \centering
  \hbox{
    \includegraphics[width=0.5\textwidth]{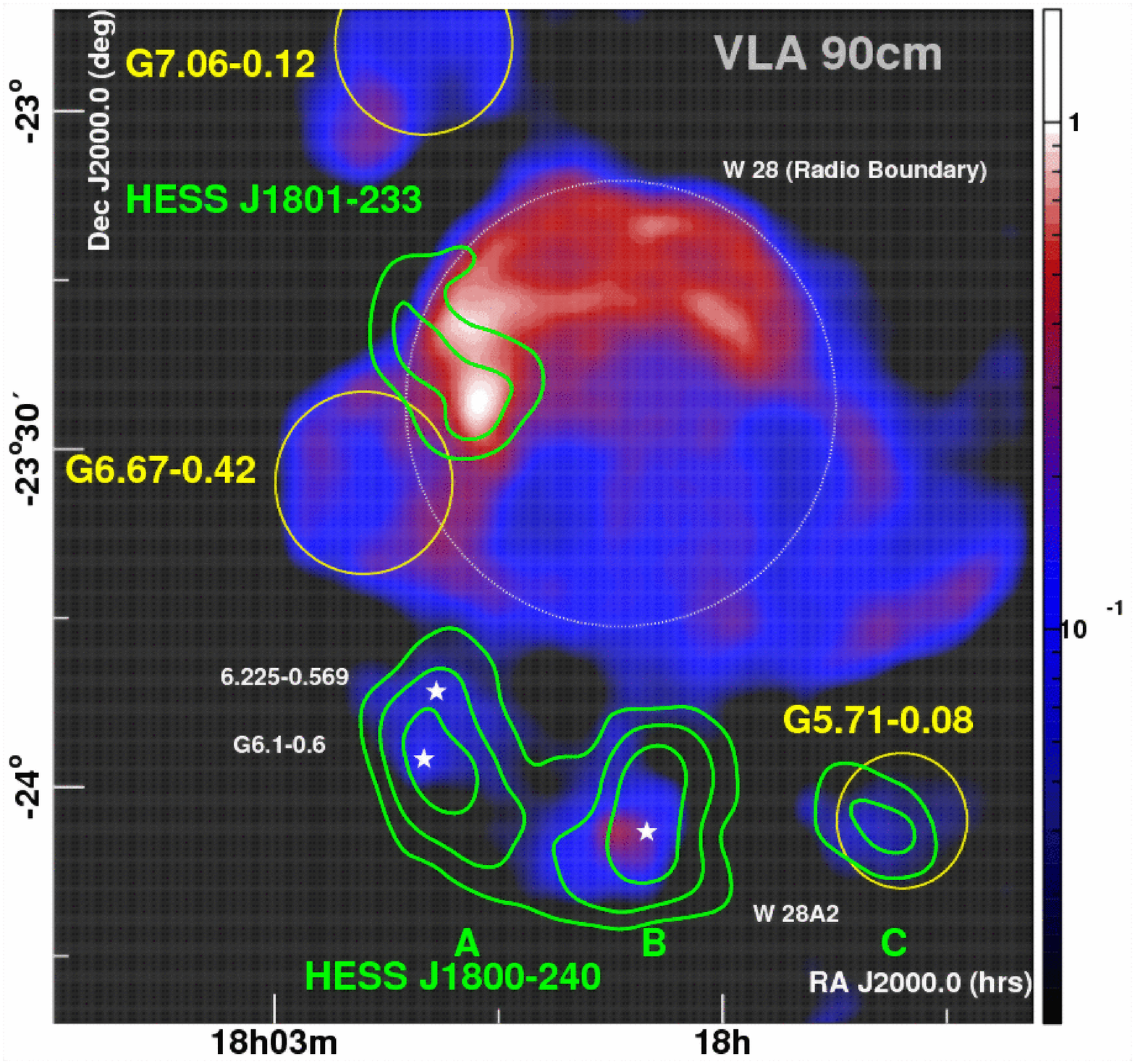}
    \includegraphics[width=0.5\textwidth]{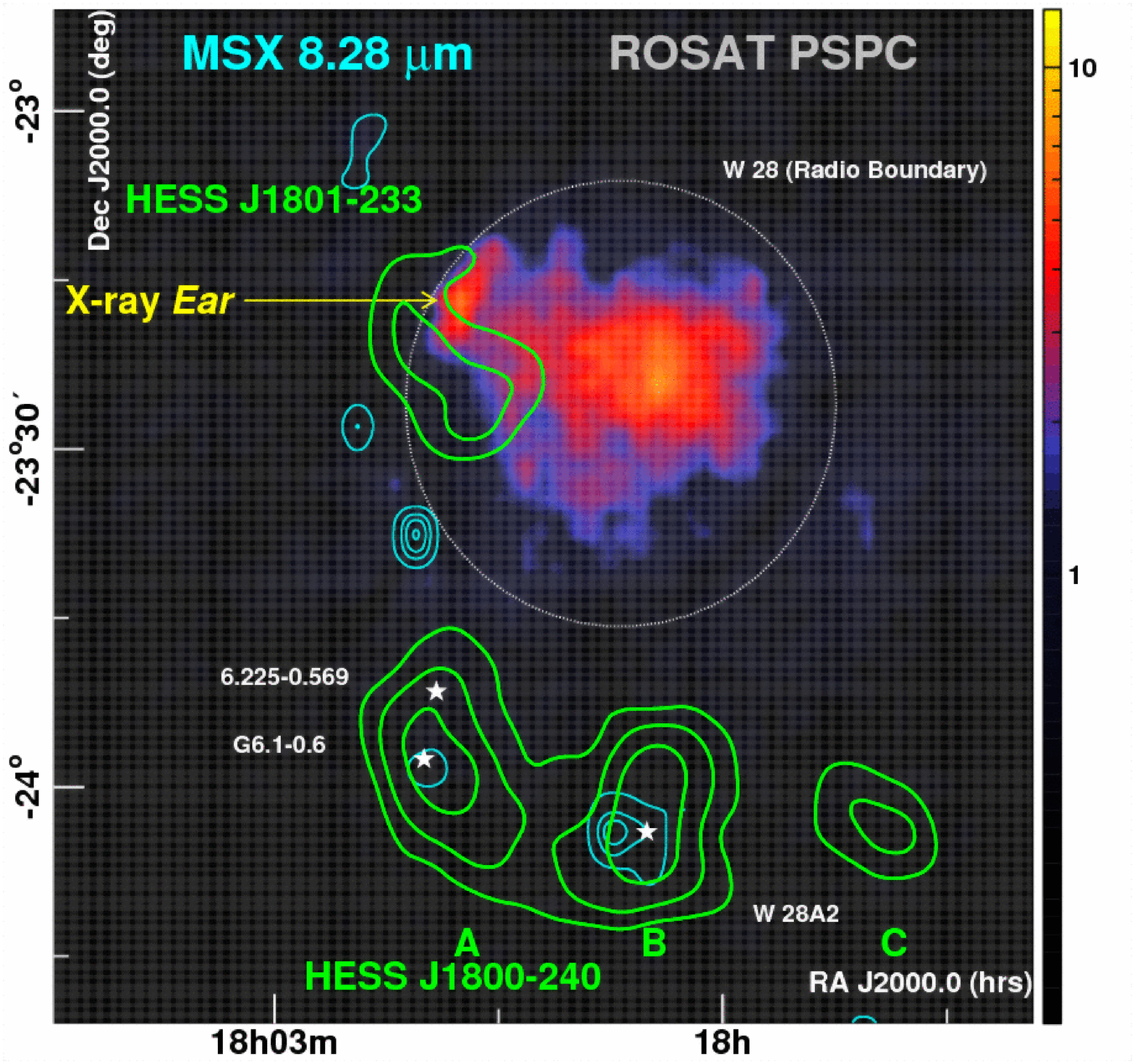}
  }
  \caption{{\bf Left:} VLA 90cm radio image from Brogan \etal (\cite{Brogan:1}) in Jy~beam$^{-1}$ 
    (rebinned by a factor 1.2 compared to the original). The VHE significance contours (green) from 
    Fig.~\ref{fig:tevskymap} are overlaid along with the HII regions (white stars) and the additional SNRs and SNR candidates 
    (with yellow circles indicating their location and approximate dimensions) discussed in the text.
    {\bf Right:} 
    ROSAT PSPC image --- 0.5 to 2.4~keV (smoothed counts per bin from Rho \& Borkowski \etal \cite{Rho:2}). Overlaid are contours 
    (cyan --- 10 linear levels up to 5$\times 10^{-4}$~W~m$^{-2}$~sr$^{-1}$) from the MSX~8.28~\micro m image. Other contours and
    objects are as for the left panel. The X-ray {\em Ear} representing a peak at the northeastern edge is indicated.}
   \label{fig:mwl}
\end{figure*}

Figure~\ref{fig:mwl} compares the radio (left panel), infrared and X-ray views (right panel) of the W~28 region with the VHE 
significance contours.
The Very Large Array (VLA) 90~cm continuum radio image from Brogan \etal (\cite{Brogan:1})
illustrates the shell-like SNR morphology peaking strongly along the northern and eastern boundaries. HESS~J1801$-$233 
can be seen to overlap the northeastern shell of the SNR, coinciding with a strong peak in the 90~cm continuum emission.
We note that a thermal component is likely present in this peak, given its spectral index $\alpha \sim -0.2$ 
(for $S \propto \nu^\alpha$) between 90 and 20~cm (Dubner \etal \cite{Dubner:1}).
Outlines of the SNRs traced by non-thermal radio emission, G6.67$-$0.42 and \object{G7.06$-$0.12} 
(Yusef-Zadeh \etal \cite{Yusef:1}, Helfand \etal \cite{Helfand:1}, 
labelled as G6.51$-$0.48 and G7.0$-$0.1 by Brogan \etal \cite{Brogan:1}) are also indicated.
In addition, Brogan \etal notes that the non-thermal radio arc \object{G5.71$-$0.08}, which overlaps well with HESS~J1800$-$240C,
could be a partial shell and therefore an SNR candidate. The distances to G6.67$-$0.42 and G5.71$-$0.08 are 
presently unknown.
Directly south of W~28, the ultracompact HII region W~28A2 is a prominent radio source, 
and is positioned within $0.1^\circ$ of the centroid of HESS~J1800$-$240B. The other HII regions 
G6.1$-$0.6 (Kuchar \& Clark \cite{Kuchar:1}) and 6.225$-$0.569 (Lockman \cite{Lockman:1}) are also associated with radio
emission.

The X-ray morphology as shown (Fig.~\ref{fig:mwl} right panel) in the ROSAT PSPC (0.5 to 2.4~keV) image 
from Rho \& Borkowski (\cite{Rho:2}) reveals the central concentration of X-ray emission, which is predominantly 
thermal in nature with characteristic temperatures in 
the range $kT \sim$0.4 to 2~keV. An X-ray peak or {\em Ear} lies
at the northeastern boundary 
and just outside the 4$\sigma$ significance contour of HESS~J1801$-$233. 
A non-thermal component to the {\em ear} emission (3$\times$1.5$^\prime$) 
(2.1$\times 10^{-14}$~erg cm$^{-2}$ s$^{-1}$ at 1~keV) with a power-law 
index $\Gamma$=1.3 has been suggested by Ueno \etal (\cite{Ueno:1})
based on XMM-Newton observations in the 0.5 to 10~keV energy range.
The total kinetic energy of the SNR is estimated at 
$\sim 4 \times 10^{50}$~erg, which could be a lower limit due to the possible break-out of the SNR along the southern edge  
away from the molecular cloud to the north and east (Rho \& Borkowski \cite{Rho:2}). 
The HII regions, W~28A2 and G6.1$-$0.6 
are prominent in the 8.28~\micro m image (Fig.~\ref{fig:mwl} right panel) from the Midcourse Space Experiment (MSX), showing that 
a high concentration of heated dust still surrounds these very young stellar objects.  

\section{Discussion}
\label{sec:discussion} 

Our discovery of VHE $\gamma$-ray emission associated with dense ($n \ge 10^3$~cm$^{-3}$) molecular clouds
in the W~28 field adds to the list of such associations after the detection of diffuse $\gamma$-ray emission 
from the Galactic Ridge (Aharonian \etal \cite{HESS-Diffuse}), the association of HESS~J1834$-$087 with the old-age SNR W~41
(Lemi\'ere \etal \cite{Lemiere:1}, Albert \etal \cite{Albert:1}) and VHE emission discovered from IC~443 (Albert \etal \cite{Albert:2}). 
The VHE/molecular cloud association could indicate a hadronic origin for the parent multi-TeV particles 
where the $\gamma$-ray emission (multi-GeV to TeV energies) arises from the decay of 
neutral pions resulting from the interaction of accelerated protons (and higher $Z$ nuclei) with ambient matter of density $n$.
In this case the $\gamma$-ray flux would scale with cloud mass or density, and the total energy in accelerated particles or
CRs penetrating the cloud(s). 
We note that a perfect correlation between the VHE and molecular cloud morphologies  
is not expected due to complex time and energy-dependent propagation of CR to and within the cloud 
(see Gabici \etal \cite{Gabici:1} for a discussion). Projection effects are also likely to be important for the examples
discussed here since the VHE emission could have contributions from clouds at different velocities, not necessarily physically
connected to one another.
For example the relationship between HESS~J1801$-$233 and the W~28/molecular cloud interaction is not entirely clear due 
to the overlapping molecular cloud components at $V_{\rm LSR}>$10~km~s$^{-1}$.

One should also consider accelerated electrons as the source of $\gamma$-ray emission, via
inverse-Compton (IC) scattering of ambient soft photon fields and/or non-thermal Bremsstrahlung from 
the interaction of electrons with dense ambient matter. 
Maximum electron energies however may be considerably lower (factor $\sim$10 or more than that of protons) due to synchrotron 
cooling in magnetic fields and low shock speeds, in the absence of strong electron replenishment.
An assessment of the role of accelerated electrons requires consideration
of the non-thermal radio and X-ray emission (where a convincing measurement of the latter is so far lacking), and also magnetic fields 
in this region. Such observations will also provide constraints on synchrotron emission expected from secondary electrons resulting
from primary hadron interactions with ambient matter (as discussed above).
Relatively high magnetic fields $B\sim 100 (n/10^4\,{\rm cm^{-3}})^{0.5}$~\micro G are inferred in 
dense molecular clouds (Crutcher \etal \cite{Crutcher:1}). In addition, higher values are indicated from Zeeman splitting 
measurements in the compact areas (arcsecond scale) surrounding the  1720~MHz OH masers of the northeastern interaction 
region (Hoffman \etal \cite{Hoffman:1}), coinciding with HESS~J1801$-$233.  
To the north of W~28, another potential source of particle acceleration is PSR~J1801$-$23, where the VHE emission may
arise in an asymmetric pulsar-wind-nebula (PWN) scenario (a primarily leptonic scenario), similar to HESS~J1825$-$137 
(Aharonian \etal \cite{hessj1825}).  
However with a spin-down power of $\dot{E} \sim 6.2\times 10^{34}$ erg~s$^{-1}$ at distance $d>9.4$~kpc, this pulsar 
appears unlikely to power any of the $\gamma$-ray sources observed in the region. A PWN scenario would therefore require
a so far undetected energetic pulsar.

\begin{table}
  \caption{Details for the molecular clouds towards the VHE sources in the W~28 field, assuming a distance $d$.}
  \label{tab:masses}
  \begin{center}
    \begin{tabular}{llllll}\\
      VHE Source        & $V_{\rm LSR}$   & $d$   & $^\dagger M$  & $^\ddagger n$  & $^\S k_{\rm CR}$ \\
      &   (km~s$^{-1}$) & (kpc) &      &      & \\ \hline \\[-2mm] 
      HESS~J1801$-$233  & 0-25  & 2.0 & 0.5 & 1.4 & 13 \\ 
      HESS~J1801$-$233  & 0-12  & 2.0 & 0.2 & 2.3 & 32 \\ 
      HESS~J1801$-$233  & 13-25 & 4.0 & 1.1 & 0.6 & 23 \\ 
      HESS~J1800$-$240  & 0-20  & 2.0 & 1.0 & 1.0 & 18 \\ 
      HESS~J1800$-$240A & 12-20 & 4.0 & 1.0 & 0.7 & 28 \\ 
      HESS~J1800$-$240B & 0-12  & 2.0 & 0.4 & 2.3 & 18 \\ 
      HESS~J1800$-$240B & 12-20 & 4.0 & 1.5 & 1.2 & 19 \\ [1mm] \hline 
      \multicolumn{6}{l}{\scriptsize $\dagger$ Cloud mass $\times 10^5\,M_\odot$}\\
      \multicolumn{6}{l}{\scriptsize $\ddagger$ Cloud density $\times 10^3$~cm$^{-3}$}\\
      \multicolumn{6}{l}{\scriptsize $\S$ Cosmic-ray density enhancement, $k_{\rm CR}$ above the local value required to}\\ 
      \multicolumn{6}{l}{\scriptsize \hspace*{1mm} produce the $E>1$~TeV VHE $\gamma$-ray emission (using Eq.10 of  Aharonian (\cite{Aharonian:1})).}
    \end{tabular}
  \end{center}
\end{table}
In the case of a hadronic origin and following Eq.10 of Aharonian (\cite{Aharonian:1}), we can estimate the
CR density enhancement factor $k_{\rm CR}$ in units of the local CR density required to explain the VHE emission, given an 
estimate for the cloud masses and assumptions on distance. 
Converting the VHE energy spectra in Table~\ref{tab:locations} to an integral value for $E>1$~TeV, assuming distances
of 2 and 4~kpc for the various cloud components, and that all the VHE emission in each source is associated with the cloud component
under consideration, we arrive at values for $k_{\rm CR}$ in the range 13 to 32 (Table~\ref{tab:masses}). 

Overall, these levels of CR enhancement factor would be expected in the neighbourhood of CR accelerators such as SNRs. 
If the clouds were all at $\sim$2~kpc, an obvious candidate for such particle acceleration is the SNR W~28, 
the most prominent SNR in the region.
Despite its old age, multi-TeV particle acceleration may still occur in W~28  (Yamazaki \etal \cite{Yamazaki:1}), with protons 
reaching energies of several 10's of TeV depending on various SNR shock parameters such as speed, size and ambient matter 
density. In addition, CRs produced at earlier
epochs have likely escaped and diffused throughout the region, a situation discussed at length in 
Aharonian \& Atoyan (\cite{Aharonian:2}). 
Aharonian \& Atoyan show for slow diffusion (diffusion coefficient at 10~GeV $D_{10}\sim$10$^{26}$~cm$^2$~s$^{-1}$ as might 
be expected in dense environments) CR enhancement factors in the required range 
could be found in the vicinity (within 30~pc -- note that if at 2~kpc distance, HESS~J1800$-$240 would lie $\sim$10~pc from the 
southern circular boundary of W~28) 
of a canonical SNR as an impulsive accelerator up to $\sim10^5$~yr after the SN explosion (see their Fig.1). 
In this sense, W~28 as a source of CRs in the region could be plausible scenario.
  
The W~28 field however is a rich star formation region, and several additional/alternative sources of CR acceleration may be 
active. The SNR G6.67$-$0.42 is positioned directly to the southeast of HESS~J1801$-$233 (Fig.~\ref{fig:mwl} left panel) while 
the SNR G7.06$-$0.12 is
situated $\sim 0.25^\circ$ north of HESS~J1801$-$233 and on the west side of the HII region M~20. M~20 itself may also be an 
energy source for the molecular clouds in this region.
The SNR candidate G5.71$-$0.08 (Brogan \etal \cite{Brogan:1}) may also be responsible in some way for HESS~J1800$-$24C given the
good positional overlap between the two. These radio SNR/SNR candidates are without a distance estimate making it 
unclear as to how they relate to the molecular clouds in the region.
The morphology of HESS~J1800$-$240 displays several peaks, perhaps resulting from changes in cloud density and/or the
presence of additional particle accelerators and local conditions. For HESS~J1800$-$24B, a potential energy source is the 
unusual ultra-compact HII region W~28A2 (G5.89$-$0.39),
representing a massive star in a very young phase of evolution. W~28A2 
exhibits very energetic bipolar molecular outflows (Harvey \& Forveille \cite{Harvey:1}, Acord \etal \cite{Acord:1},
Sollins \etal \cite{Sollins:1}) which may arise from the accretion of matter by the progenitor star. 
The outflow ages are estimated at between  $\sim 10^3$ to $10^4$~yr. Recent observations 
(Klaassen \etal \cite{Klaassen:1}) suggest both outflows extend over a combined distance of 
$\sim 2^\prime$ (or $\sim$1.2~pc at $d=2$~kpc), with total kinetic energy of $3.5\times 10^{46}$~erg.  
Surrounding the outflows is a very dense ($>10^4$~cm$^{-3}$) molecular envelope of diameter 0.5$^\prime$ to 1$^\prime$.
Despite the lack of any model to explain multi-TeV particle acceleration in such HII regions, its kinetic energy budget 
and its spatial overlap with a VHE source makes W~28A2 a tempting candidate for such acceleration.
Already, there are two examples of VHE emission possibly related to the environments of hot, young stars --- 
TeV~J2032+4130 (Aharonian \etal \cite{HEGRA_TEVJ2032}) and HESS~J1023$-$575 (Aharonian \etal \cite{HESS_WR20A}).
In this context, the HII regions G6.1$-$0.6 and 6.225$-$0.569 may also play a similar role in HESS~J1800$-$24A. 
Among the prominent open clusters in the area,  NGC~6523 and NGC~6530 $\sim 0.5^\circ$ southeast of HESS~J1800$-$240, and 
NGC~6514 associated with M~20 $\sim 0.7^\circ$ north of HESS~J1801$-$233 may also 
provide energy for CR production. Finally, if the VHE emission is associated with truly distant cloud components 
approaching the Scutum-Crux arm at $\sim$4~kpc, undetected background particle accelerators would then play a role.

\begin{figure}[t!]
  \centering
  \includegraphics[width=9.5cm]{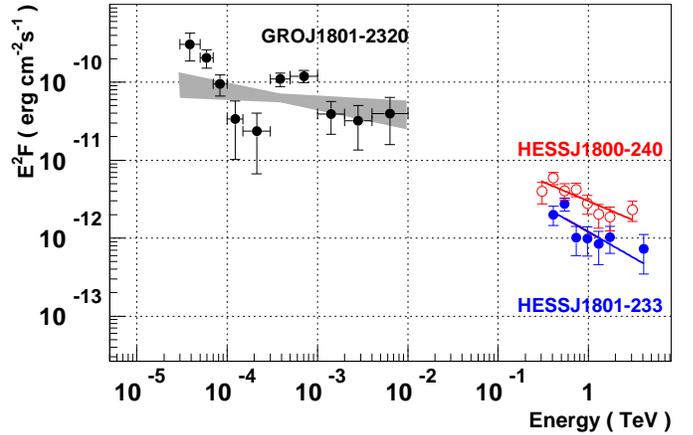}
  \caption{Energy fluxes of HESS~J1801$-$233 and HESS~J1800$-$240 (for regions 
    defined in Tab.~\ref{tab:locations}) compared to the $E>100$~MeV counterpart GRO~J1801$-$2320.
    The power law fits and data points (summarised in Tab.~\ref{tab:locations}) are also indicated: 
    HESS~J1801$-$233 (solid blue line and points); HESS~J1800$-$240 (open red points and solid line);
    GRO~J1801$-$232  (solid black points and grey 1$\sigma$ confidence band).}
  \label{fig:spectrum}
\end{figure}

Fig.~\ref{fig:spectrum} also compares the EGRET and VHE spectra. Given the degree-scale EGRET PSF, GRO~J1801$-$2320 remains
unresolved at scales of the VHE sources. Although the peak of the EGRET emission coincides with HESS~J1801$-$233, we therefore cannot rule
out unresolved MeV/GeV components from HESS~J1800$-$240. 
Observations with GLAST will be required to determine the MeV/GeV components of the VHE sources.

\section{Conclusions}
\label{sec:conclusion} 
In conclusion, our observations with the H.E.S.S. $\gamma$-ray telescopes have revealed VHE $\gamma$-ray sources
in the field of W~28 which positionally coincide well with molecular clouds. HESS~J1801$-$233 is seen toward the northeast
boundary of W~28, while HESS~J1800$-$240 situated just beyond the southern boundary of W~28 comprises three components. 
Our studies with NANTEN $^{12}$CO(J=1-0) data show molecular clouds spanning a broad range in local standard of rest velocity
$V_{\rm LSR}=$5 to $\sim$20~km~s$^{-1}$, encompassing the distance estimates for W~28 and various star formation sites in the 
region. If connected, and at a distance $\sim$2~kpc, the clouds may be part of a larger parent 
cloud possibly disrupted by W~28 and/or additional objects related to the active star formation in the region.
Cloud components up to $\sim$4~kpc distance ($V_{\rm LSR}>$10~km~s$^{-1}$) however, remain a possibility.

The VHE/molecular cloud association could indicate a hadronic origin for the VHE sources in the W~28 field.
Under assumptions of connected cloud components at a common distance of 2~kpc, or, alternatively, 
separate cloud components at
2 and 4~kpc, a hadronic origin for the VHE emission implies cosmic-ray densities $\sim$10 to $\sim$30 times the local value.
W~28 could provide such densities in the case of slow diffusion.
Additional and/or alternative particle accelerators such as HII regions representing very young stars, 
other SNRs/SNR candidates and/or several open clusters in the region may also be contributors. 
Alternatively, if cloud components at $V_{\rm LSR}>$10~km~s$^{-1}$ are at distances $d\sim4$~kpc,
as-yet undetected particle accelerators in the Scutum-Crux arm may be responsible.
Detailed modeling (beyond the scope of this paper), and further multiwavelength observations of this region
are highly recommended to assess further the relationship between the molecular gas and potential particle accelerators 
in this complex region, as well as the nature of the acclerated particles. 
In particular, further sub-mm observations (eg. at high CO transitions) will provide more accurate cloud mass estimates, and allow
to search for disrupted/shocked gas towards the southern VHE sources. Such studies
will be valuable in determining whether or not W~28 and other energetic sources have disrupted 
molecular material at line velocities $>$10~km~s$^{-1}$.

\begin{acknowledgements}
The support of the Namibian authorities and of the University of Namibia
in facilitating the construction and operation of H.E.S.S. is gratefully
acknowledged, as is the support by the German Ministry for Education and
Research (BMBF), the Max Planck Society, the French Ministry for Research,
the CNRS-IN2P3 and the Astroparticle Interdisciplinary Programme of the
CNRS, the U.K. Particle Physics and Astronomy Research Council (PPARC),
the IPNP of the Charles University, the Polish Ministry of Science and 
Higher Education, the South African Department of
Science and Technology and National Research Foundation, and by the
University of Namibia. We appreciate the excellent work of the technical
support staff in Berlin, Durham, Hamburg, Heidelberg, Palaiseau, Paris,
Saclay, and in Namibia in the construction and operation of the equipment.
The NANTEN project is financially supported from JSPS (Japan Society
for the Promotion of Science) Core-to-Core Program, MEXT Grant-in-Aid
for Scientific Research on Priority Areas, and SORST-JST (Solution
Oriented Research for Science and Technology: Japan Science and
Technology Agency).
We also thank Crystal Brogan for the VLA 90~cm image and the referee
for valuable comments.
\end{acknowledgements}

\end{document}